%% file: 3541.tex
\documentclass{aa}                        
\usepackage{graphicx}                   
\usepackage{natbib}                       
\bibpunct{(}{)}{;}{a}{}{,}               

\begin{document}

\title{A Search for {\ion{O}{vi}} in the Winds of B-Type Stars}

\author{J. Zsarg\'o\inst{1}\fnmsep\inst{2}
\and A. W. Fullerton\inst{2}\fnmsep\inst{3}
\and N. Lehner\inst{4}\fnmsep\inst{2}
\and D. Massa\inst{5}}

\offprints{J.Zsarg\'o, \email{jzsargo@astro.phyast.pitt.edu};
A.W.Fullerton, \email{awf@pha.jhu.edu};
N.Lehner, \email{nl@astro.wisc.edu}; D.Massa
\email{massa@taotaomona.gsfc.nasa.gov}}

\institute{ \emph{Present Address:} Dept. of Physics
and Astronomy,
University of Pittsburgh, 3941 O'Hara str.,
Pittsburgh, PA 15260, USA
\and Dept. of Physics and Astronomy,
The Johns Hopkins University,
3400 N. Charles St.,
Baltimore, MD 21218, USA
\and Dept. of Physics and Astronomy,
University of Victoria,
P.O. Box 3055, Victoria, BC, V8W 3P6, Canada
\and \emph{Present Address:} Dept. of Astronomy,
University of Wisconsin,
475 N. Charter St.,
Madison, WI 53706, USA
\and SGT Inc., Code 681.0,
 NASA's Goddard Space Flight Center,
Greenbelt, MD 20771, USA}

\date{Recieved 28 January 2003/ Accepted 23 April 2003}

\abstract{We have conducted a survey of {\it FUSE} spectra of 235 Galactic B-type stars
in order to determine the boundaries in the H-R diagram for the production
of the superion {\ion{O}{vi}} in their winds.
By comparing the locations and morphology of otherwise unidentified 
absorption features in the vicinity of the {\ion{O}{vi}} resonance doublet with
the bona fide wind profiles seen in archival {\it IUE} spectra of the resonance 
lines of {\ion{N}{v}}, {\ion{Si}{iv}} and {\ion{C}{iv}}, we were able to detect
blueshifted {\ion{O}{vi}} lines in the spectra of giant and supergiant
stars with temperature classes as late as B1.
No features attributable to {\ion{O}{vi}} were detected in dwarfs later than
B0, or in stars of any luminosity class later than B1, although
our ability to recognize weak absorption features in these stars is severely
restricted by blending with photospheric and interstellar features.
We discuss evidence that the ratio of the ion fractions of {\ion{O}{vi}} and
{\ion{N}{v}} is substantially different in the winds of early B-type stars
than O-type stars.
}

\maketitle

\section{Introduction}\label{section:intro}

The presence of {\ion{O}{vi}} in the stellar winds of early-type stars
was one of the early surprises of far-UV observations with the {\it Copernicus}
satellite.
It was originally difficult to understand how observable amounts of {\ion{O}{vi}},
which is two stages above the dominant form of oxygen ({\ion{O}{iv}}), exist
in the winds of these stars.
All but the very
hottest O stars emit too few photons above the ionization thresholds
for either {\ion{O}{v}} (77.4~eV) or {\ion{O}{vi}} (113.9~eV) for
photoionization by photospheric radiation to be a viable formation mechanism.
Initial attempts to explain the presence of {\ion{O}{vi}} postulated
the presence of high-temperature coronal zones near the base of the wind.
{\cite{lam76}} were able to reproduce the observed amount of {\ion{O}{vi}}
in late O-type stars with coronal models, but could not account for
its presence in the winds of stars with spectral types later than $\sim$B0.

Instead, {\citet{cas79}} noticed that the ``two stages above dominant''
character of the {\ion{O}{vi}} superion could be explained naturally
by Auger ionization if X-rays were present in the winds of hot stars.
By this mechanism, photons with $h \nu \ga$ 580~eV ($\lambda \la$
21.23~\AA)
remove an electron from the inner shell of the {\ion{O}{iv}} ion to produce
{\ion{O}{v}} in a highly excited state.  This X-ray photoionization is
followed by recombination through an Auger cascade, which results
in the ejection of an additional electron and the creation of
an {\ion{O}{vi}} ion; see \citet{ope90} for details. The predicted X-ray flux,
necessary for {\ion{O}{vi}} production, was subsequently detected by observations
from the {\it Einstein} {\citep{har79,sew79}} and {\it ROSAT}
{\citep{ber96}} observatories.
Since then, a variety of models have been proposed to explain the
origin of these X-rays, including
a deep-seated, hot corona {\citep[e.g.,][]{cas79,wal84}},
shocks due to the intrinsic line de-shadowing instability
{\citep[e.g.,][]{luc80,owo88,fel97}}, and
heating of plasma in magnetic loops {\citep[e.g.,][]{cas81,udd02}}.
Although the origin of these X-rays is still uncertain, calculations
by {\citet{mac93,mac94}} incorporating X-rays distributed
throughout the wind have demonstrated that the observed levels of superionization
can be reproduced.
As a result, Auger ionization is now understood to be responsible
for the production of superions in the winds of hot stars.

A straightforward consequence of the Auger mechanism is that, in the
presence of a sufficient flux of suitably energetic X-rays, a particular
superion will be produced only as long as its parent ion (for CNO, the
ion two stages below it; \citet{ope90}) remains abundant.
Since the dominant state of O switches from {\ion{O}{iv}} to {\ion{O}{iii}}
at {$T_{\rm eff}$} $\approx$ 30 kK, {\citet{cas79}} predicted that {\ion{O}{vi}}
would not be present in spectra of stars later than $\sim$B0.5.
This was in reasonably good agreement with the survey for wind features in
{\it Copernicus} spectra of 47 early-type stars by {\citet{sno76}},
who claimed to detect {\ion{O}{vi}} in stars as late as B1.
However, subsequent analysis of the same spectra by {\citet{mor79}} resulted
in the retraction of the detections for both B1-type stars.
Consequently, the precise boundary in the H-R diagram for the
occurrence of the superion {\ion{O}{vi}} has remained controversial.

With the launch of the {\it Far Ultraviolet Spectroscopic Explorer}
({\it FUSE}) in 1999, it has once again become possible to obtain
far-ultraviolet spectra of large numbers of early-type stars with
high spectral resolution.
Owing to its higher sensitivity, many more targets are available to {\it FUSE}
than {\it Copernicus}.
For example, after $\sim$2.5 years of operation, {\it FUSE} had obtained
more than 200 spectra of B-type stars, whereas {\it Copernicus} observed
$\sim$40 over its lifetime.
The availability of this significantly larger sample of {\it FUSE} spectra
motivated us to re-investigate the location of the temperature boundary
where the production of {\ion{O}{vi}} in the winds of B-type stars ceases.
Our survey for stellar {\ion{O}{vi}} is described in this paper, which is
organized as follows.
The observational material is described in \S\ref{section:obs},
while the techniques used to identify weak stellar {\ion{O}{vi}} absorptions
is discussed in \S\ref{section:method}.
Illustrative detections are presented in \S\ref{section:result} for each
spectral class, and the results of the survey are summarized in
\S\ref{section:sum}.

\section{Observational Material}\label{section:obs}

\subsection{The {\it FUSE} Sample of B-Type Stars}

By the autumn of 2002, {\it FUSE} observations of 235 B-type stars were available
for this survey.
Most of these spectra were obtained for programs developed by the
Principal Investigator Team, especially surveys of interstellar
{\ion{O}{vi}} in the halo and disk of the Galaxy.
These spectra were supplemented by observations collected for
various Guest Investigator programs for which the proprietary period had
elapsed.
The distribution of this sample with temperature class is
illustrated in Fig.~\ref{statistics}, in which we also coarsely
indicate the distribution with luminosity class by open and
shaded bars.
Figure~\ref{statistics} shows that the sample is strongly biased toward
luminous B0--B3 stars.
Although subject to many selection effects, the {\it FUSE} sample is
substantially larger than the {\it Copernicus} sample of 41 B-type stars
{\citep[see, e.g.,][]{sno77}} that is the basis for previous efforts
to set boundaries for the occurrence of stellar {\ion{O}{vi}} in the
H-R diagram.

\subsection{{\it FUSE} Observations and Data Reduction}

The spectroscopic instrumentation on {\it FUSE} and its inflight performance
have been described by {\citet{moo00}} and {\citet{sah00}}, respectively.
The B-star spectra used in this survey were generally obtained through the
$30\arcsec \times 30\arcsec$ (LWRS) aperture.
The fainter targets were obtained in time-tag (TTAG) mode, where the
location and arrival time of each photon is recorded, while brighter
targets were observed as spectral images in time-integrated histogram
(HIST) mode.
The observations of our sample were approximately evenly split between
these two observing modes.

The standard calibration software, {\tt CALFUSE v.2.0.5}, was used
to extract and calibrate the spectra.
For TTAG data, the photon event lists for different exposures associated
with a given observation were first concatenated, and the entire list
was used as input to {\tt CALFUSE}.
Data obtained in HIST mode were processed through {\tt CALFUSE}
exposure-by-exposure, then aligned and coadded to obtain the
observation-level spectrum.
The resultant flux-calibrated spectra are in the heliocentric reference frame,
and are characterized by spectral resolution of $\sim$20~{km\,s$^{-1}$} (FWHM)
for all instrumental configurations.
However, owing to the diverse purposes for which the spectra were
originally obtained, there is a wide range in the signal-to-noise
ratio (S/N) across the sample.
The resultant S/N of any given spectrum depends on the brightness of
the source, the total integration time devoted to it, and the extent
to which systematic motions (planned or due to thermal drifts in the
instrument) of the spectrum in the dispersion direction reduced
the fixed-pattern noise inherent to the micro-channel plate detectors.

\subsection{Ancillary Data}

Whenever possible, we retrieved archival {\it IUE} SWP spectra of our targets
from the Multimission Archive at Space Telescope (MAST) in order to compare the
morphology of stellar features at longer UV wavelengths with those accessible
to {\it FUSE}.
These data were processed with the standard {\tt NEWSIPS} software, and
were retrieved as MX browse files.
We also retrieved low resolution (FWHM $\sim$ 0.2 {\AA}) far-ultraviolet
spectra of {$\tau$} Scorpii
obtained by {\it Copernicus} {\citep{sno77}}.

\section{Methodology}\label{section:method}

Following {\citet{sno76}} and {\citet{mor79}}, we sought wind features
attributable to {\ion{O}{vi}} on a star-by-star basis by comparing the
positions and morphologies of spectral features in the vicinity of the
{\ion{O}{vi}} doublet with bona fide stellar wind features in {\it IUE} spectra
of the {\ion{Si}{iv}}, {\ion{C}{iv}}, and {\ion{N}{v}} resonance doublets.
The fundamental properties of these transitions are summarized
in Table~\ref{reslines}.
All comparisons were made in the heliocentric frame of reference,
since reliable determinations of the systemic velocity are not
available for many of the stars in the sample.
For {\ion{O}{vi}}, we relied on $FUSE$ spectra from the LiF1 channel,
which has the largest effective area in the vicinity of the {\ion{O}{vi}}
doublet.
Redundant spectra from other channels were used to verify the reality of
weak features and to ensure that fixed-pattern noise in the detector did
not cause spurious detections.

This straightforward approach is limited primarily by blending with
interstellar or stellar features, which becomes an increasingly serious
impediment to the detection of features formed in intrinsically weak stellar
winds, such as those expected for cooler and less luminous B-type stars.
Figure~\ref{illustration} shows the effects of interstellar contamination
on {\it FUSE} spectra of three early B-type stars in the region of
the {\ion{O}{vi}} doublet.
The rest wavelengths of strong interstellar features are indicated by
``combs'' that extend over the width in the upper two panels.
Most of this contamination is caused by H$_2$, which typically blankets
much of the spectral region accessible to {\it FUSE}.
However, the most serious blends affecting the detection of stellar
{\ion{O}{vi}} are due to interstellar  Lyman~${\beta}$ line of {\ion{H}{i}}
at 1026 \AA~ and a mixture of {\ion{C}{ii}}, {\ion{C}{ii}*}, and {H$_2$}
absorption near 1037 \AA.
Both features are generally strong for Galactic sight lines and can mask
wind features at heliocentric velocities $|v| \geq$ 1400 {km\,s$^{-1}$} and
$|v| \leq$ 500 {km\,s$^{-1}$} in the blue and red components of {\ion{O}{vi}},
respectively.
These blends severely compromise the detection of stellar wind absorption
features characterized by small velocities; see, e.g., \S\ref{section:B0}.

The complexity of the underlying photospheric spectrum also affects
the detectability of weak {\ion{O}{vi}} features formed in the stellar wind.
For example, the lowest panel of Fig.~\ref{illustration} shows that
photospheric line blanketing is extremely severe for giants and supergiants
later than B2; see \S\ref{section:B2+} for further discussion.
As a less severe example, Figure~\ref{model} shows synthetic spectra of an
early B-type supergiant in the vicinity of the strategic wind lines, which
was kindly computed for us by Dr. Paul Crowther with the non-LTE,
line-blanketed model atmosphere program for the expanding atmospheres of
early-type stars {\tt CMFGEN} {\citep{hil98}}.
The model corresponds approximately to a B0 supergiant with the following
parameters: {{$T_{\rm eff}$} = 25\,000~K}, {$\log L/L_{\sun}$= 5.42},
{$\dot{M}$} = 8$\times$10$^{-7}$~{${\rm M}_{\sun} \, {\rm yr}^{-1}$}, and
{${v}_\infty$} = 1\,200~{km\,s$^{-1}$}.
The synthetic spectra were rotationally broadened by {$v \sin i$}= 80~{km\,s$^{-1}$},
which is a typical value of the projected rotational velocity for the
stars presented here.
The models naturally produce P~Cygni profiles for the {\ion{C}{iv}}
and {\ion{Si}{iv}} resonance doublets, but the {\ion{N}{v}} and {\ion{O}{vi}}
superions are missing because Auger ionization via X-rays were not
incorporated.
Figure~\ref{model} shows that the region of the {\ion{O}{vi}} doublet
is the most seriously affected by photospheric lines, most of which are
due to {\ion{Fe}{iii}}.
However, it is also apparent that these lines do not mimic the spacing of
the components of the {\ion{O}{vi}} doublet, so there is little likelihood
of confusion.
Nevertheless, the presence of photospheric lines frequently impaired our
ability to perform a detailed optical depth comparison of wind features.

An alternative approach to identifying weak {\ion{O}{vi}} wind features that
are confused with stellar or interstellar lines is to search for synchronous
spectral variability at the doublet separation.
This approach is valid because no {\ion{O}{vi}} is expected in the
stellar photosphere, while interstellar lines are not expected to vary.
{\citet{leh03}} report the results of a survey for {\ion{O}{vi}} variability
based on multiple observations made by {\it FUSE}.
Even though the observing schedule was not optimized for the detection
of stellar wind variability, they found that $\sim$64\% of the stars earlier
than B1 were variable.
The variability survey includes several of the B-type targets investigated
here, and provides an opportunity to verify our technique for identifying
{\ion{O}{vi}} features based on spectral morphology alone.

\section{Results} \label{section:result}

Owing to the intrinsic weakness of their winds, most stars in our sample were
not likely to exhibit fully developed P~Cygni profiles in {\ion{O}{vi}}.
Instead, we expected to see blueshifted absorption that is typically
stronger near the blue edge of the profile (which may or may not
be the terminal velocity of wind, {${v}_\infty$}).
Although careful inspection of the {\it FUSE} B-star sample produced
many likely identifications of {\ion{O}{vi}} wind features, it was
not always possible to confirm their nature due to blending with
interstellar or photospheric lines.
In many cases, the variability study
of {\citet{leh03}} confirmedthat these features can indeed be attributed to stellar winds.
Whenever our comparison of {\ion{C}{iv}}, {\ion{Si}{iv}}, and {\ion{O}{vi}}
strongly suggested the presence of wind features in {\ion{O}{vi}
and multiple observations were available, the {\citet{leh03}} study
revealed time variability in {\ion{O}{vi}}.

The most compelling example is the case of {\object{HD 187459}} (B0.5 Ib),
where {\citet{leh03}} found variability at wind velocities
between $-750$~{km\,s$^{-1}$} and $-$1550~{km\,s$^{-1}$}.
Figure~\ref{HD187459} shows the observed {\ion{O}{vi}}, {\ion{N}{v}},
{\ion{Si}{iv}}, and {\ion{C}{iv}} profiles for {\object{HD 187459}}.
The broad feature in {\ion{O}{vi}} around $-$1100~{km\,s$^{-1}$} was
suspected to be blueshifted wind absorption that is also present
in the {\ion{Si}{iv}} and {\ion{C}{iv}} profiles.
Although the overlap between the interstellar {\ion{H}{i}} L$_{\beta}$
line and the {\ion{O}{vi}} 1031.93 {\AA} wind absorption prevented us from
unambiguously identifying stellar {\ion{O}{vi}} in {\object{HD 187459}},
the variability occurring at the relevant velocities supported our initial
suspicion.

Even though the variability study supported our conclusions in the
more clear-cut cases, it also revealed the limitations of our method.
Often variability occurred in {\ion{O}{vi}} when no wind features
could be identified by our profile comparison.
It is likely, therefore, that our detection rate seriously underestimates
the true frequency of the occurrence of {\ion{O}{vi}} wind features in
the spectra of B-type stars.
Table~\ref{result_table} illustrates this deficiency by comparing our
detection rate with those of {\citet{leh03}} for Galactic B stars.
Table~\ref{result_table} also highlights the rather small number of
B-type stars with multiple $FUSE$ observations.
Follow-up observations or time-variability studies are required to
corroborate our potential detections and determine the true frequency of
stellar {\ion{O}{vi}} in the B spectral class.

In the remainder of this section, we present the results of the profile
comparisons for stars that did provide clear evidence
for blueshifted wind features in {\ion{O}{vi}} (including 3 objects with spectral
types as late as B1) and discuss the implications of non-detections in
the spectral class B2 and beyond.
The fundamental properties of these objects and the details of the
{\it FUSE} observations are listed in Table~\ref{tab1}.
We describe these detections in order of spectral type.

\subsection{B0 and B0.5 Stars \label{section:B0}}

The presence of {\ion{O}{vi}} wind absorption in the spectra of B0 dwarfs
has been known since early observations with {\it Copernicus\/}
{\citep{lam78}}.
Detections in the fainter objects accessible to {\it FUSE} are not trivial,
because the winds of B0~V stars are usually slow
{\citep[{${v}_\infty$} $\la$ 600 {km\,s$^{-1}$} ;][]{pri90}}
and the {\ion{O}{vi}} $\lambda$1037.62 profile is almost always obscured by
strong interstellar \ion{C}{ii}, \ion{C}{ii}*, and H$_2$ absorption.
These problems are illustrated in Figure~\ref{B0V}, which compares {\it FUSE} and
{\it Copernicus} spectra of the {\ion{O}{vi}} doublet for three B0~V stars.
A broad wind profile is clearly recognizable in both {\ion{O}{vi}} lines
in the {\it Copernicus} spectrum of the nearby, lightly reddened star
{\object{$\tau$ Scorpii}}, but the saturated interstellar lines mask any
possible {\ion{O}{vi}} $\lambda$1037.62 feature in {\object{HD 97471}} and
{\object{HD 207538}}. Accordingly, we used {\object{$\tau$ Sco}} as a template
to look for strong wind features in {\ion{O}{vi} $\lambda$1031.93}.
For example, the spectrum of {\object{HD 97471}} in Fig.~\ref{B0V} exhibits a
prominent wind profile around 1032 \AA.
Remarkably, these absorptions are even stronger than the
much-studied {\ion{O}{vi}} lines of {\object{$\tau$ Sco}}.
In contrast, it is apparent from Figure~\ref{B0V} that {\object{HD 207538}}
shows no morphological evidence for the presence of wind absorption at
{\ion{O}{vi}}.
By using this method, we detected prominent {\ion{O}{vi}} wind profiles in the
spectra of $\sim$19\% of the 31 B0~IV--V stars (see Table~\ref{result_table}),
which emphasizes the importance of superionization in the winds of B0 dwarfs.

Since the terminal velocities of B0 giants and supergiants are usually
larger than 1000 {km\,s$^{-1}$} , wind absorption in the red component of the
{\ion{O}{vi}} doublet is less likely to be masked by interstellar
contamination.
About a third of the luminous B0--B0.5 stars with {\it FUSE} spectra suggested
the presence of {\ion{O}{vi}} wind absorption.
Of course, this fraction represents a lower limit on the true frequency of
occurrence, since blending with interstellar and stellar features
is still a problem; and indeed, not all the stars with suspected
{\ion{O}{vi}} absorption can be claimed as firm detections at this point.
{\object{HD 219188}} is clear-cut case, which is discussed in detail below.

\subsubsection{\object{HD 219188}}

{\object{HD 219188}} (B0.5 II--IIIn) is runaway star that has normal elemental
abundances {\citep{con90}}, appears to be single {\citep{gie86,phi96}},
and evidently formed far from the Galactic plane {\citep{kee86}}.
The morphology of its {\it FUSE} spectrum in the vicinity of {\ion{O}{vi}}
is compared with the well-developed P~Cygni profiles in the
resonance lines of {\ion{Si}{iv}} and {\ion{C}{iv}} in Fig.~\ref{HD219188}.
Figure~\ref{HD219188} shows a strong, narrow blueshifted absorption at
about $-$1300 {km\,s$^{-1}$} ($\sim$0.9\,{${v}_\infty$}) that occurs in both components of the
{\ion{O}{vi}} doublet.
The location and width of these absorptions correspond to similar
features in the {\ion{Si}{iv}} and {\ion{C}{iv}} wind profiles.
Consequently, we interpret the blueshifted {\ion{O}{vi}} absorptions as wind
features that are formed predominantly in the high-velocity portion
of the outflow.

{\object{HD 219188}} also offers a nice opportunity to determine the optical
depths of
the blueshifted absorption features in the spectra of the {\ion{O}{vi}} doublet.
In contrast to most other targets, the effective continuum in the vicinity of
these features can be placed reliably because the large projected
rotational velocity ({$v \sin i$} = 197~{km\,s$^{-1}$}) broadens and smooths
its underlying photospheric spectrum.
Figure~\ref{ratios} shows the measured optical depths for each component
in 20~{km\,s$^{-1}$} wide velocity bins (upper panel) and the derived ratio
of the optical depth in the blue and red components (lower panel).
Within the uncertainties, the optical depth ratio is 2, which is consistent
with expectations for an optically thin plasma.
Thus, the position, morphology, and optical depth ratio of the absorption
components are consistent with the interpretation that they arise from
optically thin {\ion{O}{vi}} near the terminal velocity of the stellar wind.

The measured optical depth of the high-velocity component in
{\ion{O}{vi}} $\lambda$1032, together with the absence of analogous
features in the superion of {\ion{N}{v}} (Fig.~\ref{HD219188})
implies that the relative ion fractions of {\ion{O}{vi}} and {\ion{N}{v}}
are unusual.
At any wind velocity, the radial Sobolev optical depth of a wind profile is
$\tau_{rad} \propto f\,\lambda_0\,q_i\,A_E\,\dot{M}\,( {\rm d}v / {\rm d}r )^{-1}$,
where $f$, $\lambda_0$, $q_i$, and $A_E$ are respectively the oscillator
strength, rest wavelength, ion fraction, and elemental abundance associated
with  a particular transition.
For solar abundances and atomic parameters for the blue component
of the {\ion{O}{vi}} and {\ion{N}{v}} resonance doublets, this implies
$q_i$({\ion{O}{vi}})/$q_i$({\ion{N}{v}}) $= 0.176\,\tau_{rad}$({\ion{O}{vi}})/
$\tau_{rad}$({\ion{N}{v}}).
With $\tau_{rad}$({\ion{O}{vi}}) $\approx$ 1.4 (Fig.~\ref{ratios}) and
assuming an upper limit on the non-detection of
$\tau_{rad}$({\ion{N}{v}}) $\le$ 0.1, there results
$q_i$({\ion{O}{vi}})/$q_i$({\ion{N}{v}}) $\ga$ 2.5.
In contrast, {\citet{mas03}} derived values of  0.1 -- 1.7 for this
ratio from their study of {\it FUSE} spectra of O-type stars in the
Large Magellanic Cloud, with (mean, median) values of (0.44, 0.33).
Thus, fractionally more {\ion{O}{vi}} and/or less {\ion{N}{v}} appears to be
produced in the wind of {\object{HD 219188}} than is typically produced in
the winds of O-type stars.

\subsection{B1 Stars \label{section:B1}}

As discussed in \S\ref{section:intro}, the detection of {\ion{O}{vi}} wind features
in the spectra of B1-type stars is a crucial test of theoretical predictions
of the production of high ions via Auger ionization.
Only 16 stars bridging this interface are contained in
the catalog of {\it Copernicus} observations published by {\citet{sno77}},
and the persistence of {\ion{O}{vi}} has been discussed only for subsets of
10 and 2 stars by {\citet{sno76}} and {\citet{mor79}}, respectively.
The report of {\ion{O}{vi}} wind absorption in spectra of
{\object{$\rho$ Leonis}} (B1 Iab) and {\object{$\gamma$ Arae}} (B1 Ib)
was retracted by
{\citet{mor79}}, with the result that the temperature boundary for
the occurrence of stellar {\ion{O}{vi}} is still uncertain.

In contrast, the {\it FUSE} sample consists of 66 B1 stars of all
luminosity classes.
A preliminary time-variability study of a subset of this sample
{\citep{leh01}} has already reported the presence of {\ion{O}{vi}} ``discrete
absorption components'' (DACs) in the spectra of {\object{HD 91597}} (B1~IIIne).
Our inspection of the full sample revealed that wind {\ion{O}{vi}}
is not uncommon in the spectra of B1 giants and supergiants.
We detected possible blueshifted absorption in the spectra of $\sim$20\% of
the B1 giants and supergiants, but did not find any evidence for wind
features in the spectra of subgiants and dwarfs (see Table~\ref{result_table}).
This might be a selection effect, since the winds of lower-luminosity
stars are characterized by much weaker signatures, even in dominant ions.
For the three stars discussed below, we detected strong absorption features
that can be identified with blueshifted {\ion{O}{vi}}.
Together with {\object{HD 91597}} {\citep{leh01}}, these objects provide the best
evidence that the superion {\ion{O}{vi}}
persists in the winds of stars with spectral types as late as B1.

\subsubsection{{\object{HD 93840}} (BN1 Ib)}

{\object{HD 93840}} is a well-known supergiant that exhibits anomalous
abundances due
to the presence of material processed through the CNO-cycle in its
atmosphere and wind {\citep{sav87,wal90,mas91}}.
Since the CNO cycle only alters the relative abundances of CNO nuclei,
the substantial overabundance of N and under-abundance of C implies
that O should also be deficient compared with solar values.
Figure~\ref{HD93840} compares the UV resonance doublets of
{\ion{O}{vi}}, {\ion{N}{v}}, {\ion{Si}{iv}}, and {\ion{C}{iv}}.
The anomalous nitrogen and carbon abundances are apparent in the strong
{\ion{N}{v}} and weak {\ion{C}{iv}} wind profiles, both of which are
unusual for the spectral type.
Blueshifted absorption is visible between $-$800 and $-$1200~{km\,s$^{-1}$}
in both components of the {\ion{O}{vi}}, {\ion{N}{v}}, and {\ion{Si}{iv}}
doublets, and is present more weakly in the {\ion{C}{iv}} doublet.
Even though O is deficient and N is enhanced, the optical depths
of the wind absorptions are similar, which again implies that
$q_i$({\ion{O}{vi}})/$q_i$({\ion{N}{v}}) is unusually large.

In contrast to the fully-developed P~Cygni profiles at {\ion{N}{v}}
and {\ion{Si}{iv}}, Fig.~\ref{HD93840} implies that the {\ion{O}{vi}} feature
probably exists only in absorption, since the flux peak near $+$200~{km\,s$^{-1}$}
is consistent with the continuum level.
It may be that most of the {\ion{O}{vi}} wind absorption occurs in strong,
comparatively narrow components near the terminal velocity.
As originally recognized by {\citet{mas91}}, similar DACs are
visible in the {\ion{Si}{iv}} lines of {\object{HD 93840}}.
DACs are commonly observed in wind profiles of early-type stars
{\citet[see, e.g.,][]{pri02}}, and are interpreted as signatures of
persistent, large-scale structures in the stellar wind.
Since most models for the formation of DACs imply the presence of
strong shocks, it would be very interesting to determine whether
{\ion{O}{vi}} is preferentially formed in the vicinity of these structures.
See {\citet{leh03}} for further evidence that {\ion{O}{vi}} absorption
is related to DACs.

\subsubsection{{\object{HD 191877}} $=$ {\object{HR 7716}} (B1 Ib)}

{\object{HD 191877}} is a supergiant with normal elemental abundances
{\citep{wol88}}.
Figure~\ref{HD191877} shows that its {\it FUSE}~spectrum has deep absorption
features with the separation of the components of the {\ion{O}{vi}} doublet,
blueshifted to approximately $-$1000~{km\,s$^{-1}$}.
The positions of these features coincides with excess absorption
in the P~Cygni profiles of the {\ion{Si}{iv}} and {\ion{C}{iv}} doublets.
As with {\object{HD 219188}}, little or no wind absorption is apparent
at the
position of the {\ion{N}{v}} doublet, which once again suggests that
$q_i$({\ion{O}{vi}})/$q_i$({\ion{N}{v}}) is larger than normally observed
in the winds of O-type stars.

Since {\object{HD 191877}} is one of the early B-type stars that was observed
multiple
times by {\it FUSE}, spectral variability can be used to disentangle
features formed in the wind from photospheric and interstellar blends.
Unfortunately, the time span between the observations was long (420 days)
and the quality of the two observations was quite different.
{\citet{leh03}} classified it as variable, possibly a member of a binary
system, and noted that the spectrum was too noisy to determine the
range of variability in velocity reliably.
Nevertheless, this cross-check corroborates the interpretation
of the absorptions as wind features.

\subsubsection{{\object{HD 215733}} (B1 II)}

{\object{HD 215733}} is a B1~II star located approximately 1.5~kpc below
the Galactic plane.
Although {\citet{wal76}} suggested that it was slightly deficient in N,
subsequent analysis by {\citet{kee82}} indicated that it has normal
abundances.
Figure~\ref{HD215733} shows very strong blueshifted absorption features
extending between $-$800 and $-$1200~{km\,s$^{-1}$} in both components of the
{\ion{O}{vi}} doublet.
Their positions coincide approximately with the positions of narrow
absorption components in the resonance doublets of both {\ion{Si}{iv}} and
{\ion{C}{iv}}, which otherwise exhibit well-developed wind profiles
with little redshifted emission.
This coincidence again suggests that {\ion{O}{vi}} is formed
preferentially at the larger velocities associated with these structures.
As with {\object{HD 191877}}, there is little or no wind absorption visible
in {\ion{N}{v}}.

\subsection{B2 and Later-Type Stars \label{section:B2+}}

We did not detect absorption features that could be identified
with {\ion{O}{vi}} for any objects with spectral types of B2 or later.
Furthermore, no time variability attributable to {\ion{O}{vi}} was observed
for such stars {\citep{leh03}}, though only seven were observed
more than once (see Table~\ref{result_table}).
Thus, from a purely observational perspective, we can now say definitively
that B1 is the latest spectral type at which {\ion{O}{vi}}
wind features are detected.

However, even if {\ion{O}{vi}} persists in the winds of cooler stars, the
selection effects discussed in \S\ref{section:method} work against its
detection.
Figure~\ref{HD92964} shows the regions around the standard selection of
resonance lines in the spectrum of {\object{HD~92964}} (B2.5~Iae;
see also Fig.~\ref{illustration}).
Although the wind of this supergiant has enough optical depth in the
resonance lines of abundant species like {\ion{C}{iv}} and {\ion{Si}{iv}}
to produce P~Cygni profiles, it is futile to search for absorption
features attributable to {\ion{O}{vi}} due to the combined effects of
heavy photospheric line blanketing and diminishing far-ultraviolet
flux levels.
Even without the devastating effects of line blanketing, the smaller terminal
velocities associated with the winds of these cooler stars complicate
the detection of weak absorption features, since the red component of the
{\ion{O}{vi}} doublet would typically be blended with interstellar lines
and could not be used to search for correlated absorption or variability
at the appropriate
doublet separation.
Thus, even if {\ion{O}{vi}} were present in the winds of these
cooler stars, it is doubtful that it could be detected.

\section{Summary}\label{section:sum}

We have conducted a comprehensive search for stellar wind features in the
resonance lines of the {\ion{O}{vi}} superion in {\it FUSE} spectra of
235 Galactic B stars.
This data set is substantially larger and more diverse than the sample
obtained by {\it Copernicus}, but suffers from greater confusion due
to blending with interstellar lines because the stars are systematically
farther away.
Although this confusion complicated the detection of weak stellar wind
features in {\ion{O}{vi}}, we found that:
\begin{itemize}

  \item Blueshifted wind absorption attributable to {\ion{O}{vi}}
        occurs in the spectra of B0 stars of all luminosity classes.
	The strength of the absorption is sometimes greater than that
	seen in {\object{$\tau$ Sco}}, which is frequently viewed as
	an extreme
or anomalous object {\citep[see, e.g.,][]{coh03}}.
	Instead, it seems that there is a class of B0 dwarfs that are similar
	to {\object{$\tau$ Sco}}, which is just the nearest and brightest
	representative.

  \item {\ion{O}{vi}} wind features persist to spectral types as late
        as B1, but only for more luminous stars (giants, bright giants, or
	supergiants).
	Since the winds of these objects have systematically greater
	optical depth than those of dwarfs, this luminosity dependence
        may just be an observational selection effect.

  \item No {\ion{O}{vi}} wind absorption was detected in spectra later
        than B1 for any luminosity class.
	Although this may very well be the physically-relevant boundary
	of the occurrence of the {\ion{O}{vi}} superion in the H-R diagram,
	the detection of weak absorption features in these spectra is
	problematic due to blending with interstellar and photospheric lines.

  \item The absorption features in {\ion{O}{vi}} are strongest near the
        terminal velocity of the wind, which suggests that superionization
	might preferentially occur at higher velocities in the winds
	of B-type stars.
	{\citet{leh03}} reached a similar conclusion based on the
	characteristics of the variability they observed.

  \item The similarities in the appearance of the {\ion{O}{vi}} wind
        absorptions in the spectra of the N-rich B1 supergiant
        {\object{HD 93840}}
	and the chemically normal B1 supergiant {\object{HD 191877}} indicate
	that the process responsible for the formation of the superion
	does not depend strongly on chemical composition.
	Irrespective of abundance anomalies, it appears that the ratio of
	ion fractions $q_i$({\ion{O}{vi}})/$q_i$({\ion{N}{v}}) is
	systematically higher in early B-type stars than in O-type stars.

\end{itemize}

These findings dispel lingering doubts over whether {\ion{O}{vi}}
persists in the winds of stars as cool as temperature class B1.
They are consistent with expectations based on the Auger mechanism
proposed by {\citet{cas79}}, though unfortunately the cool boundary cannot
be determined as rigorously as desired due to blending.
Since lightly-reddened, nearby B-type stars are generally too bright
to be observed by {\it FUSE}, it is unlikely that this issue will be resolved
in the foreseeable future.
A more fruitful avenue for further research will be to determine whether
the enhanced production of {\ion{O}{vi}} superions relative to {\ion{N}{v}}
superions in the winds of B-type stars is caused by the transition in X-ray
emission properties that occurs at $\sim$B1--B1.5 {\citep{cas94,coh97}},
but adequate X-ray measurements are necessary to resolve this question.

\begin{acknowledgements}

This work is based on observations made with the NASA-CNES-CSA
{\it Far Ultraviolet Spectroscopic Explorer}. {\it FUSE} is operated
for NASA by the Johns Hopkins University under NASA contract NAS5-32985.
We are grateful to Paul Crowther for sending us spectra of early B-type
supergiants computed with {\tt CMFGEN}.

\end{acknowledgements}


\clearpage
\begin{figure}
\resizebox{\hsize}{!}{

  \includegraphics{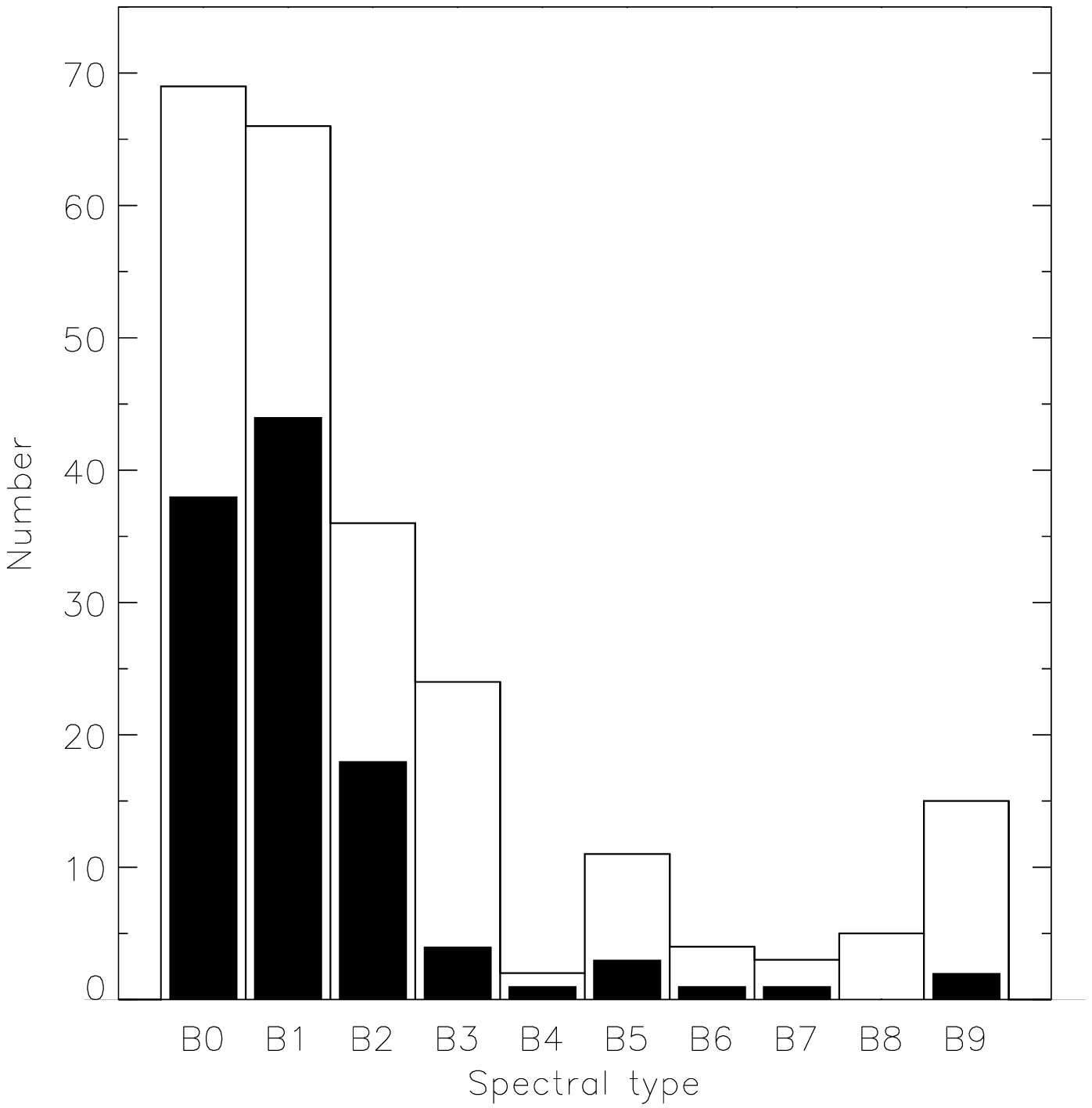}
  }
\caption{
   The distribution of Galactic B stars included in the {\it FUSE} survey
   for stellar {\ion{O}{vi}} with spectral type.
   Shaded regions in each histogram bar indicate the number of
   stars with luminosity classes I--III.
   The sample is strongly biased toward stars earlier than B3.
   }
\label{statistics}

\end{figure}

\clearpage
\begin{figure*}
\centering
\includegraphics[width=17cm]{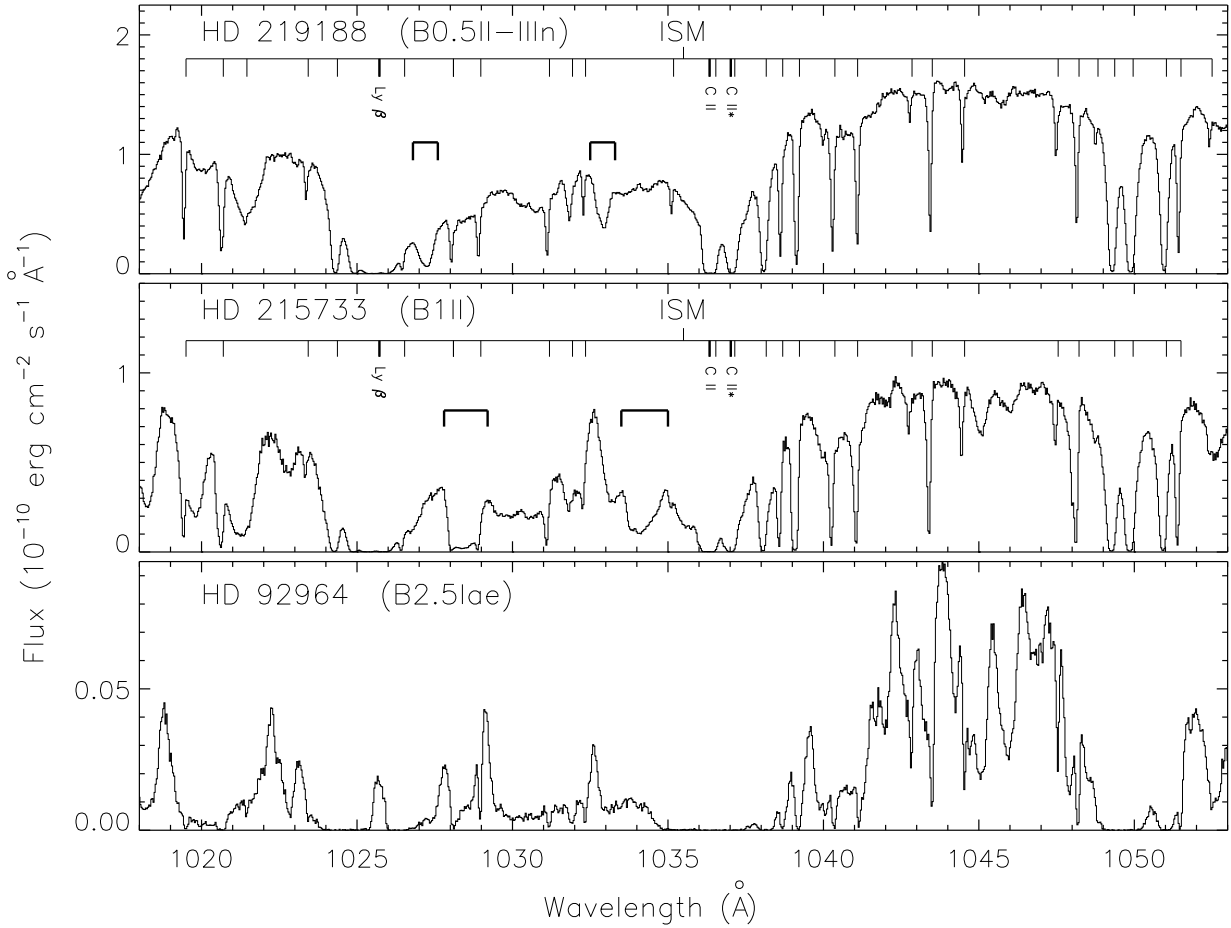}
\caption{
    The observed fluxes around the {\ion{O}{vi}} doublet toward three
    early-B stars.
    Combs in the upper two panels mark the position of interstellar
    lines, most of which are due to H$_2$.
    We identified the Lyman $\beta$ of {\ion{H}{i}} at 1026~{\AA}
    and the {\ion{C}{ii}},  {\ion{C}{ii}}* absorption complex near 1037~{\AA}
    that were particularly problematic ''contaminants''  of the wind
     {\ion{O}{vi}} absorption.
    Blue-shifted {\ion{O}{vi}} absorption features are
    indicated for {\object{HD 219188}} and {\object{HD 215733}}
    (upper two panels).
    Similar features could not be identified for  {\object{HD 92964}} (lower panel)
    due to strong photospheric line blanketing.
    }
\label{illustration}

\end{figure*}

\clearpage
\begin{figure*}
\centering
\includegraphics[width=17cm]{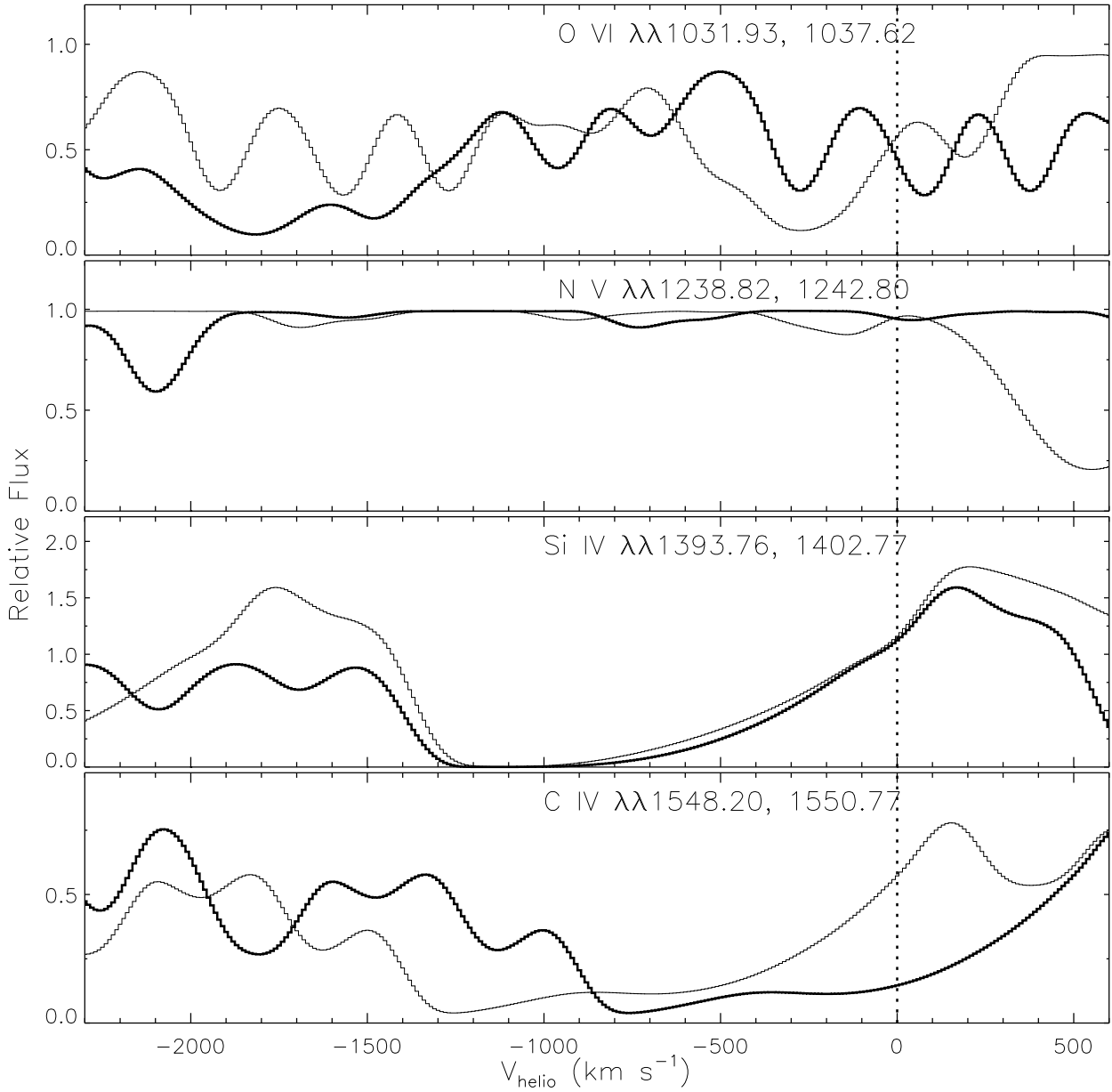}
\caption{
   Synthetic spectra between $-$2300 and $+$600~{km\,s$^{-1}$} of
   the rest wavelengths of the resonance doublets of
   {\ion{O}{vi}}, {\ion{N}{v}}, {\ion{Si}{iv}}, and {\ion{C}{iv}}
   (from top to bottom).
   The computed flux is plotted as a function of velocity  with respect to
   the laboratory wavelength of the stronger (thick line) and  weaker (thin
   line) member of each doublet.
   See \S\ref{section:method} for a description of the model parameters.
   Since the computation did not include Auger ionization via X-rays,
   {\ion{O}{vi}} and {\ion{N}{v}} were not produced.
   The photospheric features in the vicinity of {\ion{O}{vi}} are mostly
   due to {\ion{Fe}{iii}}.
   }
\label{model}
\end{figure*}

\clearpage
\begin{figure*}
\centering
\includegraphics[width=17cm]{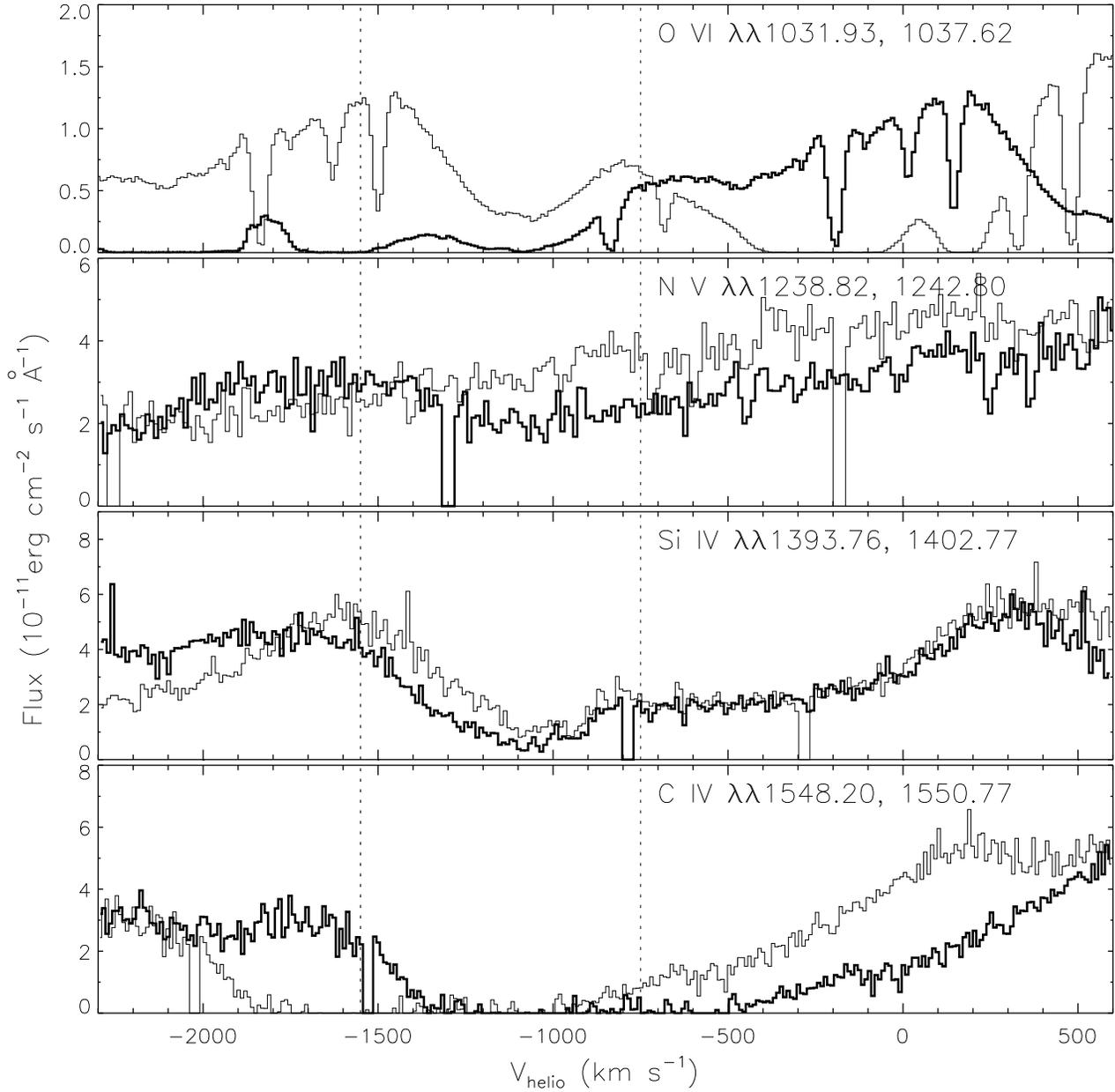}
\caption{
    Comparison of the observed wind profiles of {\object{HD 187459}} for
    the resonance
    lines of {\ion{O}{vi}}, {\ion{N}{v}}, {\ion{Si}{iv}}, and {\ion{C}{iv}}
    (from top to bottom).
    The flux is plotted as a function of velocity  with respect to the
    laboratory wavelength of the stronger (thick line) and  weaker (thin
    line) member of each doublet.
    The velocity range where {\citet{leh03}} observed time variability
    is indicated by vertical, dotted lines.
    }
\label{HD187459}
\end{figure*}

\clearpage
\begin{figure*}
\centering
\includegraphics[width=17cm]{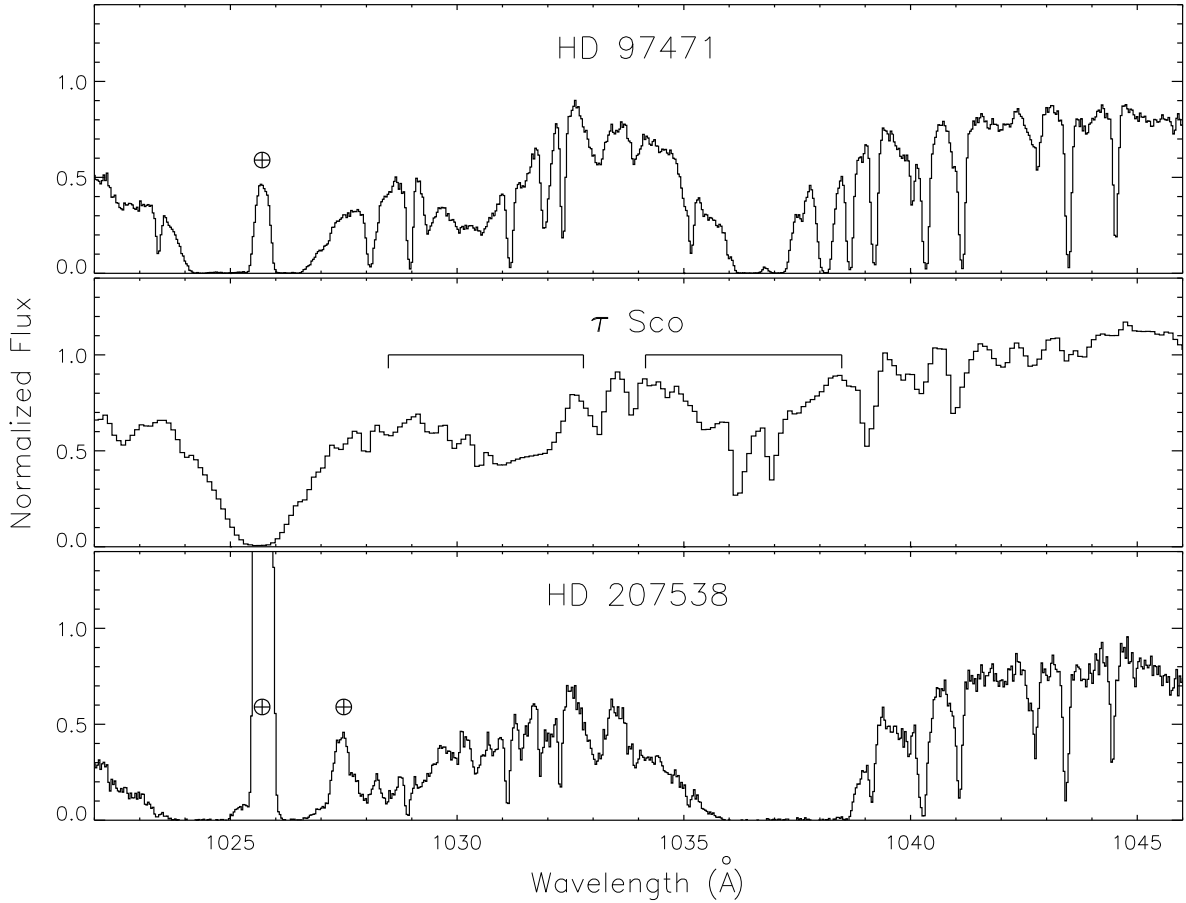}
\caption{
    Observed fluxes around the {\ion{O}{vi}} doublet for three B0~V stars.
    {\object{HD 97471}} and {\object{HD 207538}} were observed by {\it FUSE},
    while the spectrum of {\object{$\tau$ Sco}} was obtained by
    {\it Copernicus}. The positions of terrestrial airglow lines are indicated by
    the ``Earth'' symbol in the $FUSE$ spectra.
    The {\ion{O}{vi}} wind features identified by {\citet{lam78}}
    are shown above the spectrum of {\object{$\tau$ Sco}}.
    Similar strong features are apparent in the spectrum of {\object{HD 97471}},
    despite the strong interstellar {\ion{C}{ii}}, {\ion{C}{ii}*}, and
    H$_2$ lines around 1037 \AA .
    However, no {\ion{O}{vi}} wind features can be identified in the spectrum
    of {\object{HD 207538}}.
    }
\label{B0V}
\end{figure*}

\clearpage
\begin{figure*}
\centering
\includegraphics[width=17cm]{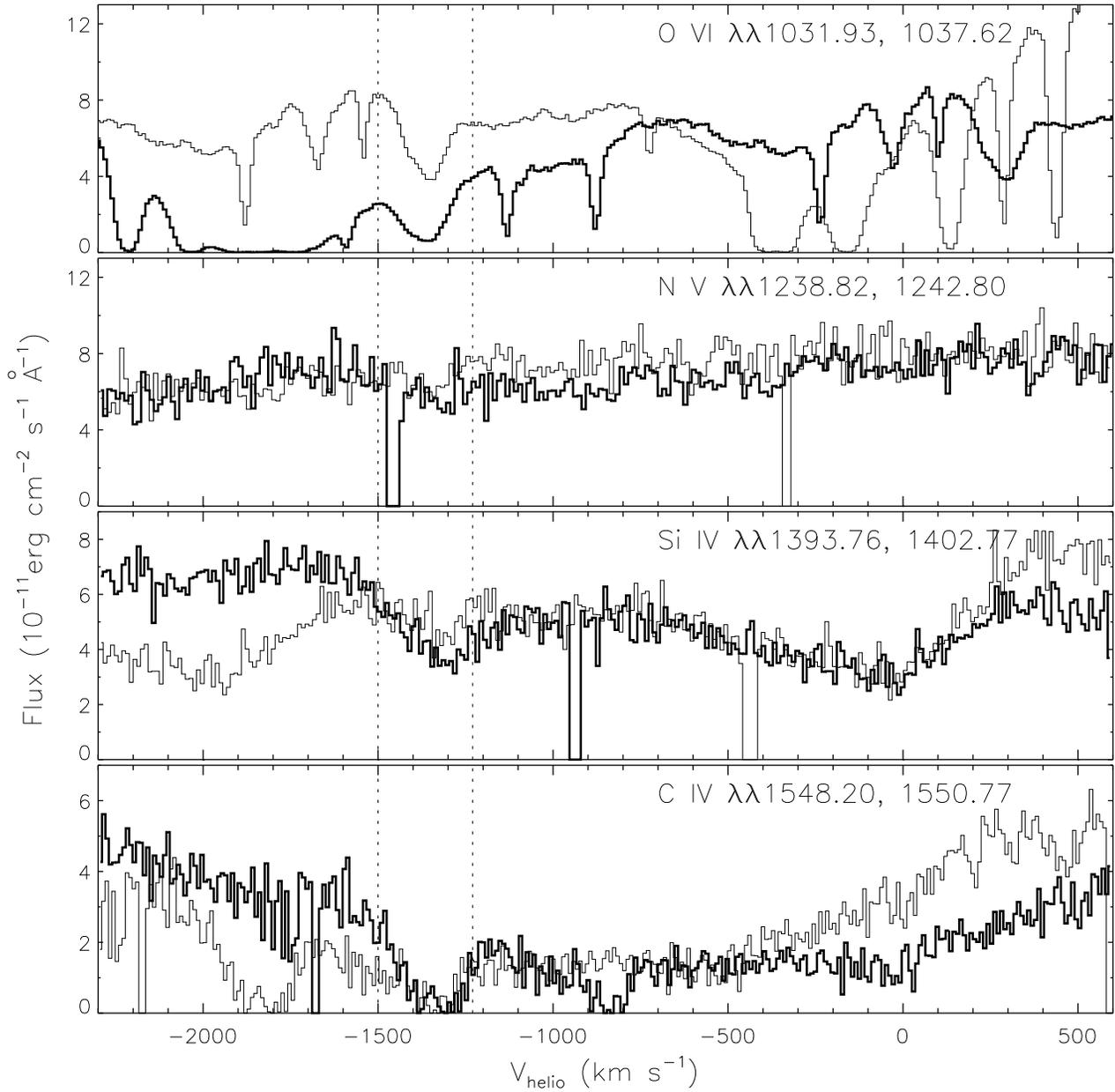}
\caption{
    Comparison of the observed wind profiles of {\object{HD 219188}} for
    the resonance
    lines of {\ion{O}{vi}}, {\ion{N}{v}}, {\ion{Si}{iv}}, and {\ion{C}{iv}}
    (from top to bottom).
    The flux is plotted as a function of velocity  with respect to the
    laboratory wavelength of the stronger (thick line) and  weaker (thin
    line) member of each doublet.
    The velocity range containing the {\ion{O}{vi}}, {\ion{Si}{iv}}, and
    {\ion{C}{iv}} wind features is indicated by vertical, dotted lines.
    Note the serious contamination of the {\ion{O}{vi}} doublet by
    interstellar lines.
    }
\label{HD219188}
\end{figure*}

\clearpage
\begin{figure}
\resizebox{\hsize}{!}{
  \includegraphics{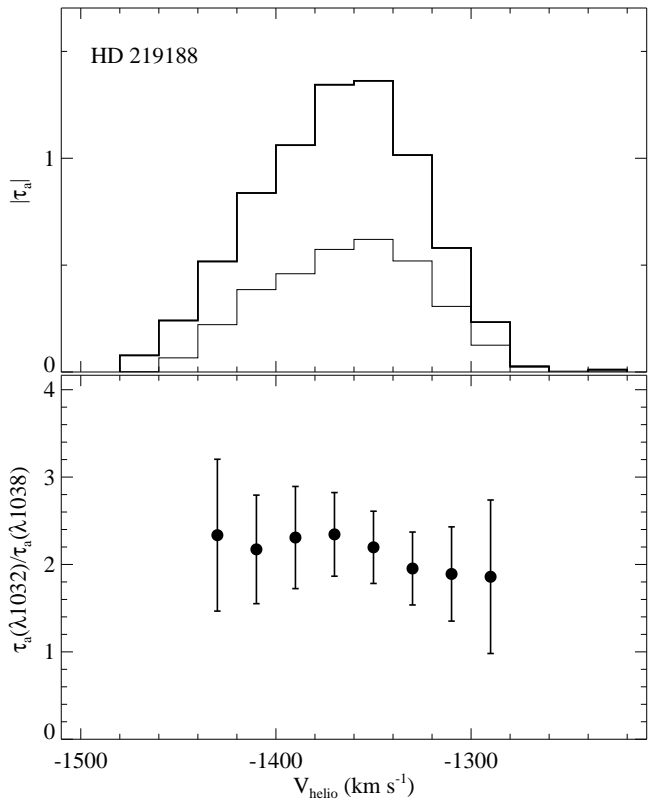}
}
\caption{
   Top panel: The measured optical depths of the blueshifted wind absorptions
   in the spectra of {\object{HD 219188}}, rebinned to 20 {km\,s$^{-1}$} velocity bins.
   The thick and thin lines show {\ion{O}{vi}} $\lambda$1031.93 and
   $\lambda$1037.62, respectively.
   Lower panel: The optical depth ratio of the \ion{O}{vi} doublet in each
   velocity bins.
   The error bars reflect the estimated uncertainties introduced by the
   continuum placement around the wind features.
   }
\label{ratios}
\end{figure}

\clearpage
\begin{figure*}
\centering
\includegraphics[width=17cm]{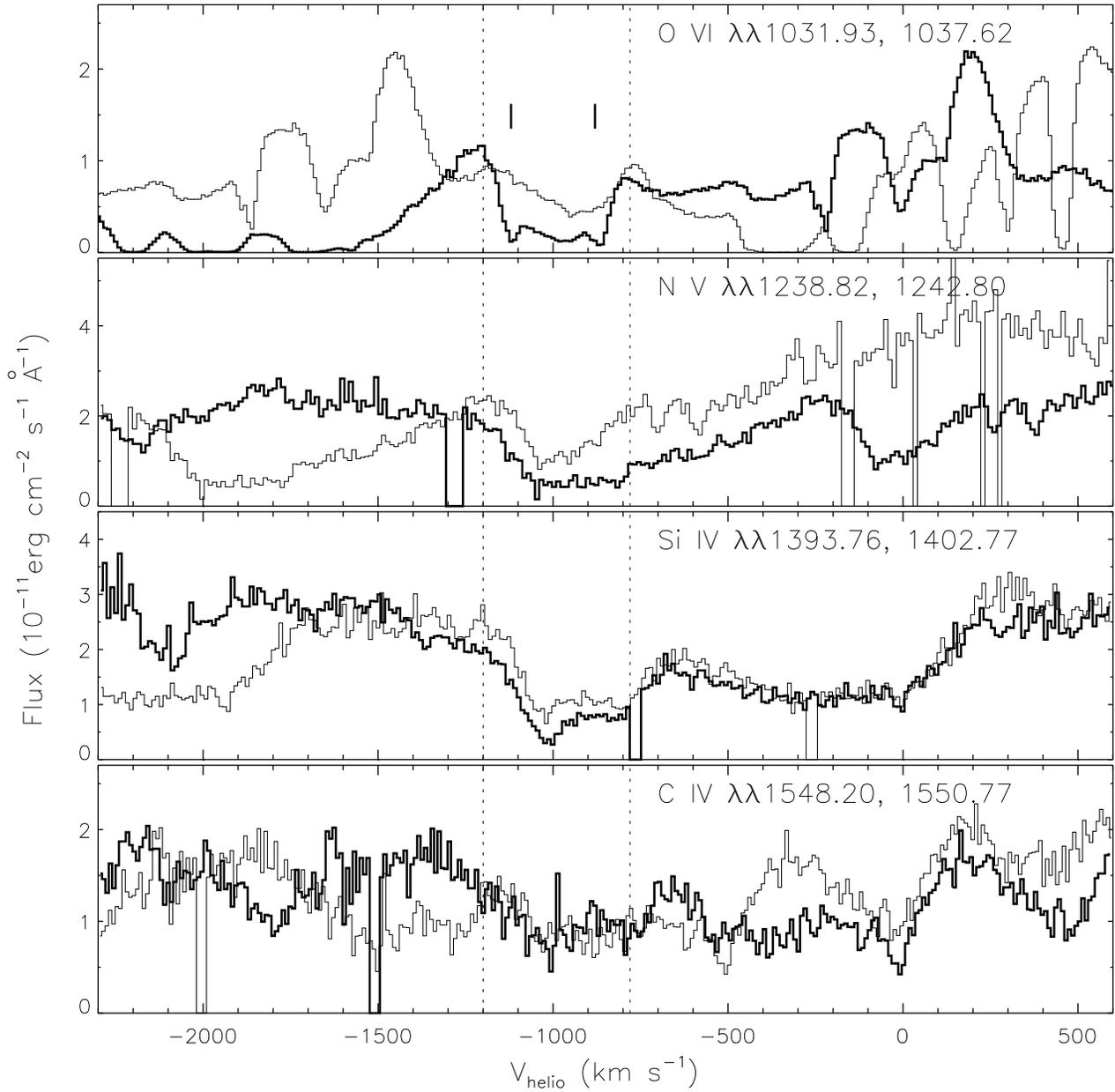}
\caption{
   Same as Figure~\ref{HD219188} for {\object{HD 93840}}.
   Interstellar lines that contaminate the {\ion{O}{vi}} absorption 
   are indicated by tick marks.
   }
\label{HD93840}
\end{figure*}

\clearpage
\begin{figure*} 
\centering
\includegraphics[width=17cm]{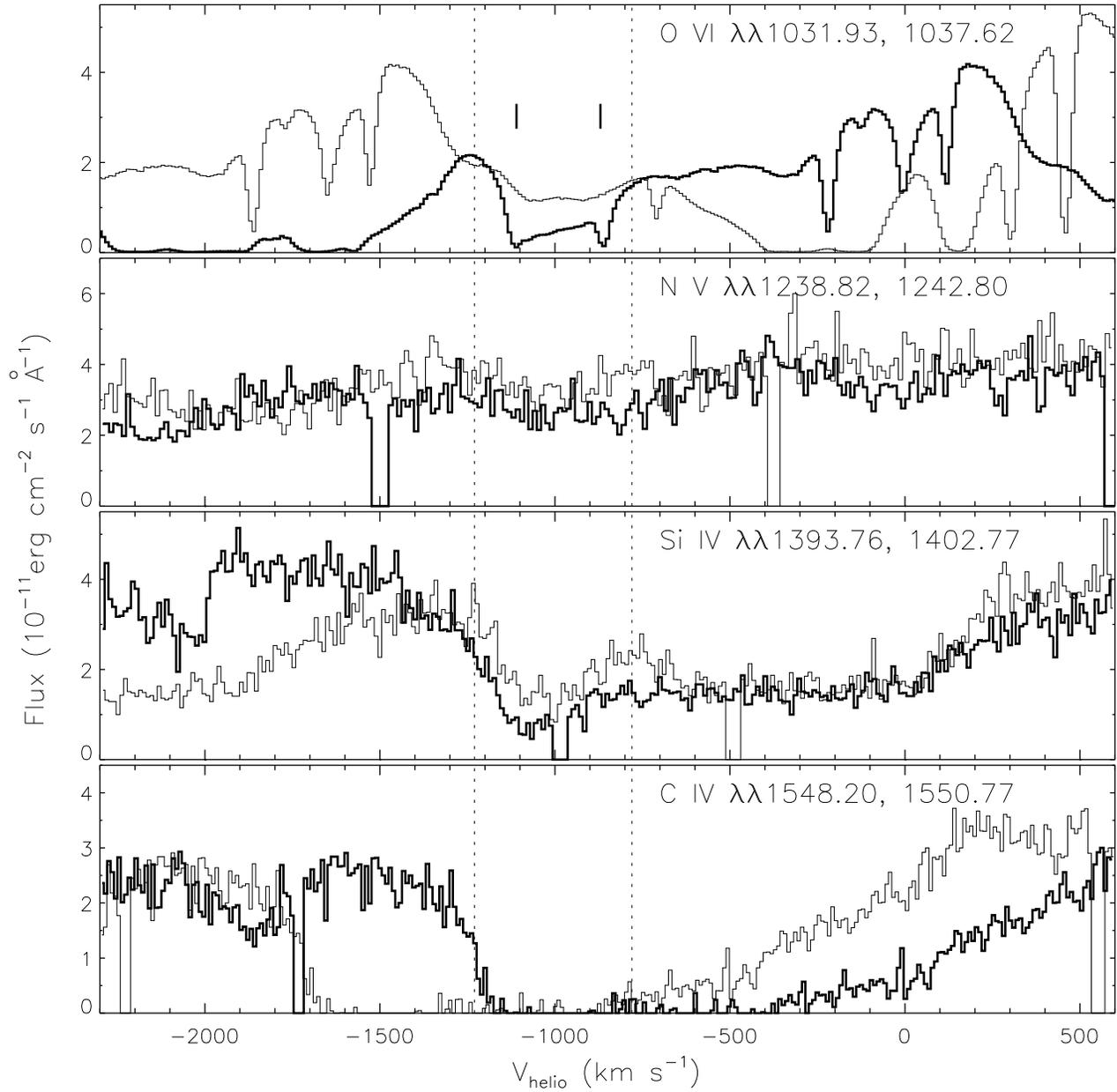}
\caption{
   Same as Figure~\ref{HD93840} for {\object{HD 191877}}. 
   }
\label{HD191877}
\end{figure*}

\clearpage
\begin{figure*} 
\centering
\includegraphics[width=17cm]{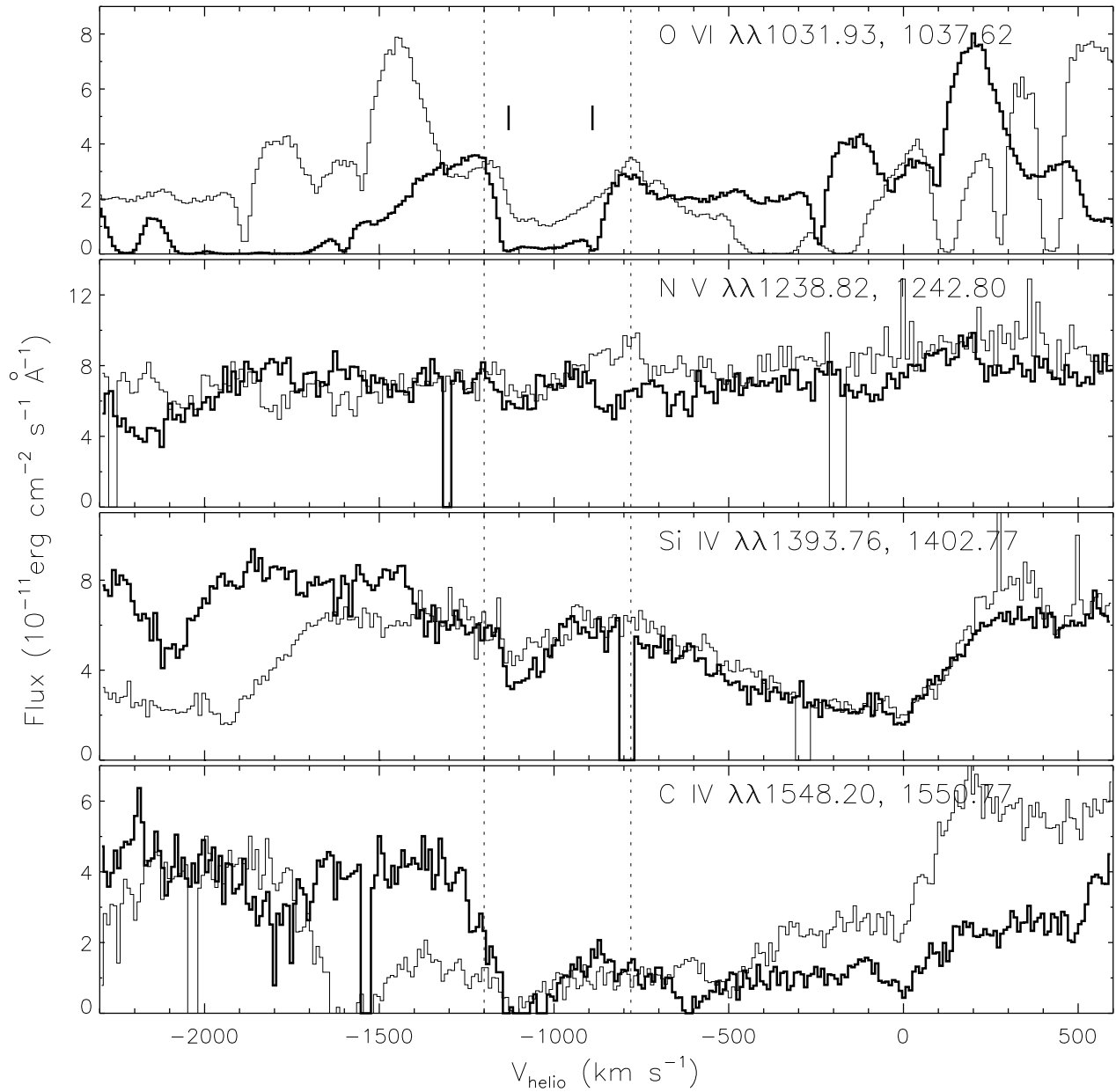}
\caption{
    Same as Figure~\ref{HD93840} for {\object{HD 215733}}.
    } 
\label{HD215733}
\end{figure*}

\clearpage
\begin{figure*}
\centering
\includegraphics[width=17cm]{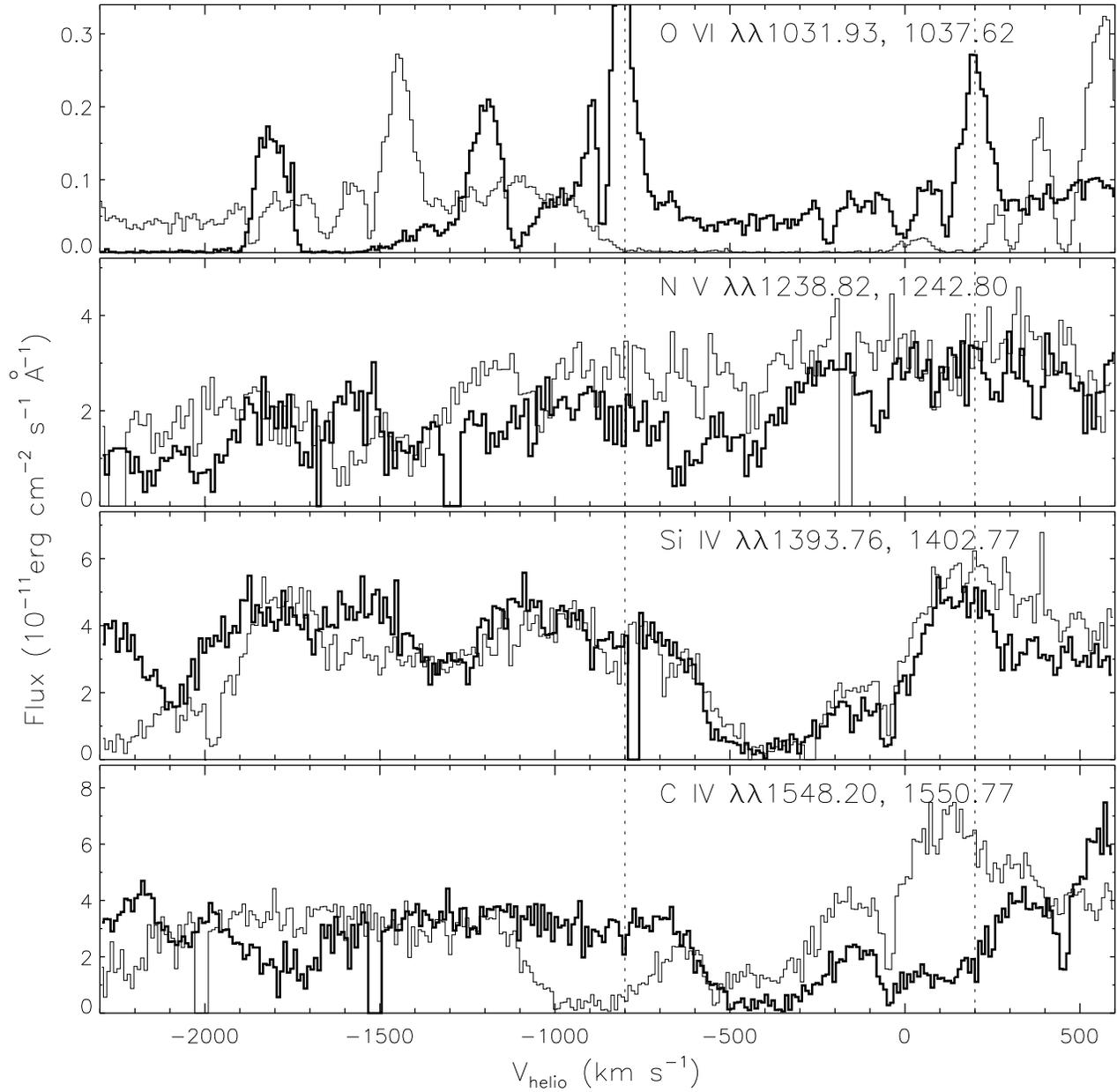}
\caption{
    Same as Figure~\ref{HD219188} for {\object{HD 92964}}.
The dotted     vertical
    lines indicate the velocity region where wind signatures are
    expected to appear.
    }
\label{HD92964}

\end{figure*}

\clearpage
\include{3541tab1}

\clearpage
\include{3541tab2}

\clearpage
\include{3541tab3}

\end{document}

%% file: 3541tab1.tex
\begin{table}
\caption{Properties of Ultraviolet Resonance Lines with Wind Profiles\label{reslines}}
\begin{tabular}{lllll} \hline\hline
Ion         &
Abundance\,$^a$ &
I.P. Range\,$^b$ &
Wavelengths\,$^c$ &
$\log f\lambda \, ^c$ \\
             &
     &
    (eV)         &
      (\AA)      &
      \\ \hline{\ion{O}{vi}}   & 8.87 & 113.9\,--\,138.1 & 1031.93, 1037.62 & 2.137, 1.836 \\ 
{\ion{N}{v}}   & 7.97 &  77.5\,--\,97.9  & 1238.82, 1242.80 & 2.289, 1.988 \\
{\ion{Si}{iv}}  & 7.55 &  33.5\,--\,45.1  & 1393.76, 1402.77 & 2.855, 2.554 \\
{\ion{C}{iv}}   & 8.55 &  47.9\,--\,64.5  & 1548.20, 1550.77 & 2.470, 2.169 \\ \hline
\end{tabular}

$^a$ -- Solar-system abundances from {\cite{gre93}} and {\citet{and89}},
        expressed as $\log N / N_H  + 12.00$.

$^b$ -- The ionization potentials are from {\cite{moo70}}.

$^c$ -- The rest wavelengths and oscillator strengths for the blue and red
        components of the doublets are from {\citet{mor91,mor02}}.
\end{table}

%% file: 3541tab2.tex
\begin{table}
\caption{The Frequency of Galactic B-type Stars with {\ion{O}{vi}}
Wind Features \label{result_table}}
\begin{tabular}{lllll} \hline\hline
Spectral Type Bin   &
&
   \multicolumn{3}{c}{ \%~(Det./Tot.)$^a$ } \\ \cline{3-5}                     &
& Zs03\,$^b$ &
& L03\,$^c$ \\ \hline B0 -- B1 (I-III)    & & $\ga$34 (13/38) & &  $\ga$86 (6/7) \\ 
B0 -- B1 (IV-V)     & & $\ga$19 (6/31)  & &  $\ga$56 (5/9) \\ 
B1 -- B2 (I-III)    & & $\ga$20 (9/44)  & &  $\ga$67 (4/6) \\ 
B1 -- B2 (IV-V)     & &  0      (0/22)  & &  0       (0/3) \\
B2 -- B9 (I-V)      & &  0     (0/100)  & &  0       (0/4) \\ \hline 
\end{tabular}

$^a$ -- Det.: The number of stars with detections; Tot.: Total number of 
objects surveyed.

$^b$ -- Present study. Includes cases with unambigous and very likely 
detections.

$^c$ -- Variability study of {\cite{leh03}}. The statistics are for Galactic
stars only.
\end{table}

%% file: 3541tab3.tex
\scriptsize
\begin{table}
\caption{Galactic B stars with {\ion{O}{vi}} Wind Features\label{tab1}}
\begin{tabular}{lllllllllll} \hline
\hline
ID &
Sp. Type & {$v \sin i$}$^a$ &
{${v}_\infty$}$^a$ &
$E$(B-V) & &{\it FUSE} Program &
Aperture/Mode$^b$ &
Integration &
{\it IUE} Data \\
   &
         & 	 ({km\,s$^{-1}$})   &	({km\,s$^{-1}$})  &	     & &     ID         &
                  &
 Time (s)   &
   Set      \\ \hline
{\object{HD 97471}} & B0~V      & $\ldots$ & $\ldots$ & $\ldots$ 
& & A1180404       & LWRS/HIST &  4200 & $\ldots$  \\ \\
{\object{HD 219188}}  & B0.5~II-I & 197     & $\ldots$ & 0.14    
& & P1018902       & MDRS/HIST &  1660 & SWP05653 \\ \\ 
{\object{HD 93840}}  & BN1~Ib    &  95     & 1235    & 0.14    
& & P1012701       & LWRS/HIST &  5318 & SWP21525 \\ \\
{\object{HD 191877}}  & B1~Ib     & 152     & 1160    & 0.18    
& & P1028701       & LWRS/HIST &  6132 & SWP14825 \\
        &           &         &         &         
& & P2051101,02,03 & LWRS/HIST & 28055 &         \\ \\
{\object{HD 215733}}  & B1~II     &  84     & 1240    & 0.11    
& & P1018602       & MDRS/HIST &  1932 & SWP07356 \\ \hline
\end{tabular}

$^a$ -- \citet{how97}.

$^b$ -- LWRS: $30\arcsec \times 30\arcsec$; MDRS: $4\arcsec  \times 20\arcsec$;
        HIST: histogram mode.
\end{table}
\normalsize

%% file: 3541.bbl
\begin{thebibliography}{}

\bibitem[Anders \& Grevesse(1989)]{and89}
Anders, E. \& Grevesse, N. 1989, Geochim. Cosmochim. Acta, 53, 197

\bibitem[Bergh\"{o}fer {et al}.(1996)]{ber96}
Bergh\"{o}fer, T. W., Schmitt, J. H. M. M., \& Cassinelli, J. P. 1996,
\aaps, 118, 481

\bibitem[Cassinelli \& Olson(1979)]{cas79}
Cassinelli, J. P. \& Olson, G. L. 1979, \apj, 229, 304

\bibitem[Cassinelli {et al.}(1981)]{cas81}
Cassinelli, J. P., Waldron, W. L., Sanders, W. T., Harnden, F. R.,
Rosner, R., \& Vaiana, G. S. 1981, \apj, 250, 677

\bibitem[Cassinelli {et al.}(1994)]{cas94}
Cassinelli, J. P., Cohen, D. H., MacFarlane, J. J., Sanders, W. T., \&
Welsh, B. Y. 1994, \apj, 421, 705

\bibitem[Cohen {et al.}(1997)]{coh97}
Cohen, D. H., Cassinelli, J. P., \& MacFarlane, J. J. 1997, \apj, 487, 867

\bibitem[Cohen {et al.}(2003)]{coh03}
Cohen, D. H., de Messi{\'e}res, G. E., MacFarlane, J. J.,
Miller, N. A., Cassinelli, J. P., Owocki, S. P., \& Liedahl 2003,
\apj, 586, 495

\bibitem[Conlon {et al.}(1990)]{con90}
Conlon, E. S., Dufton, P. L., Keenan, F. P.,
\& Leonard, P. J. T. 1990, \aap, 236, 357

\bibitem[Feldmeier {et al.}(1997)]{fel97}
Feldmeier, A., Puls, J., \& Pauldrach, A. W. A. 1997, \aap, 322, 878

\bibitem[Gies \& Bolton(1986)]{gie86}
Gies, D. R. \& Bolton, C. T. 1986, \apjs, 61, 419

\bibitem[Grevesse \& Noels(1993)]{gre93}
Grevesse, N. \& Noels, A. 1993, in Origin of the Elements, ed.
N. Prantzos, E. Vangioni-Flam, \& M. Cass\'{e}
(Cambridge: Cambridge University Press), p. 15

\bibitem[Harnden {et al}.(1979)]{har79}
Harnden, F. R., et al. 1979, \apj, 234, L51

\bibitem[Hillier \& Miller(1998)]{hil98}
Hillier, D. J. \& Miller, D. L. 1998, \apj, 496, 407

\bibitem[Howarth {et al.}(1997)]{how97}
Howarth, I. D., Seibert, K. W., Hussain, G. A. J., \& Prinja, R. K. 1997,
\mnras, 284, 265

\bibitem[Keenan {et al.}(1982)]{kee82}
Keenan, F. P., Dufton, P. L., \& McKeith, C. D. 1982, \mnras, 200, 673

\bibitem[Keenan {et al.}(1986)]{kee86}
Keenan, F. P., Brown, P. J. F., \& Lennon, D. J. 1986, \aap, 155, 333

\bibitem[Lamers \& Morton(1976)]{lam76}
Lamers, H. J. G. L. M. \& Morton, D. C., 1976, \apjs, 32, 715

\bibitem[Lamers \& Rogerson(1978)]{lam78}
Lamers, H. J. G. L. M., \& Rogerson, J. B. 1978, \aap, 66, 417

\bibitem[Lehner {et al.}(2001)]{leh01}
Lehner N., Fullerton, A. W., Sembach, K. R., Massa, D., \& Jenkins, E. B.
2001, \apj, 556, L103

\bibitem[Lehner {et al.}(2003)]{leh03}
Lehner N., Fullerton, A. W., Massa, D., Sembach, K. R., \& Zsarg\'{o}, J.
2003, \apj, in press

\bibitem[Lucy \& White(1980)]{luc80}
Lucy, L. B. \& White, R. L. 1980, \apj, 241, 300

\bibitem[MacFarlane {et al.}(1994)]{mac94}
MacFarlane, J. J., Cohen, D. H., \& Wang, P. 1994, \apj, 437, 351

\bibitem[MacFarlane {et al.}(1993)]{mac93}
MacFarlane, J. J., Waldron, W. L., Corcoran, M. F., Wolff, M. J.,
Wang, P., \& Cassinelli, J. P. 1993, \apj, 419, 813

\bibitem[Massa {et al.}(1991)]{mas91}
Massa, D., Altner, B., Wynne, D., \& Lamers, H. J. G. L. M.
1991, \aap, 242, 188

\bibitem[Massa {et al.}(2003)]{mas03}
Massa, D., Fullerton, A. W., Sonneborn, G., \& Hutchings, J. B. 2003,
\apj, 586, in press

\bibitem[Moore(1970)]{moo70}
Moore, C. E. 1970, Ionization Potentials and Ionization Limits Derived from
the Analyses of Optical Spectra (Report NSRDS-NBS34, Washington,
D.C.: US Department of Commerce)

\bibitem[Moos {et al.}(2000)]{moo00}
Moos, H. W., et al. 2000, \apj, 538, L1

\bibitem[Morton(1979)]{mor79}
Morton, D. C. 1979, \mnras, 189, 57

\bibitem[Morton(1991)]{mor91}
Morton, D. C. 1991, \apjs, 77, 119

\bibitem[Morton(2002)]{mor02}
Morton, D. C. 2002, \apjs, in preparation

\bibitem[Opendak(1990)]{ope90}
Opendak, M. G. 1990, \apss, 165, 9

\bibitem[Owocki {et al.}(1988)]{owo88}
Owocki, S. P., Castor, J. I. \& Rybicki, G. B. 1988, \apj, 335, 914

\bibitem[Philp {et al.}(1996)]{phi96}
Philp, C. J., Evans, C. R., Leonard, P. J. T., \& Frail, D. A. 1996,
\aj, 111, 1220

\bibitem[Prinja {et al.}(1990)]{pri90}
Prinja, R. K., Barlow, M. J., \& Howarth, I. D. 1990, ApJ, 361, 607

\bibitem[Prinja {et al.}(2002)]{pri02}
Prinja, R. K., Massa, D., \& Fullerton, A. W. 2002, \aap, 388, 587

\bibitem[{Sahnow} {et~al.}(2000)]{sah00}
Sahnow, D., et al. 2000, \apj, 538, L7

\bibitem[Savage \& Massa(1987)]{sav87}
Savage, B. D. \& Massa D. 1987, \apj, 314, 380

\bibitem[Seward {et al.}(1979)]{sew79}
Seward, F. D., Forman, W. R., Giacconi, R., Griffiths, R. E.,
Harnden, F. R., Jones, C., \& Pey, J. P. 1979, \apj, 234, L55

\bibitem[Snow \& Morton(1976)]{sno76}
Snow, T. P. \& Morton, D. C. 1976, \apjs, 32, 429

\bibitem[Snow \& Jenkins(1977)]{sno77}
Snow, T. P. \& Jenkins, E. B. 1977, \apjs, 33, 269

\bibitem[{{ud-Doula} \& {Owocki}(2002)}]{udd02}
{ud-Doula}, A. \& {Owocki}, S.~P. 2002, \apj, 576, 413

\bibitem[Walborn {et al.}(1990)]{wal90}
Walborn, N. R., Fitzpatrick, E. L., \& Nichols-Bohlin, J. 1990, \pasp, 102, 543

\bibitem[Walborn(1976)]{wal76}
Walborn, N. R. 1976, \apj, 205, 419

\bibitem[{{Waldron}(1984)}]{wal84}
{Waldron}, W.~L. 1984, \apj, 282, 256

\bibitem[Wollaert {et al.}(1988)]{wol88}
Wollaert, J. P. M., Lamers, H. J. G. L. M.,\& de Jager, C. 1988, \aap, 194, 197

\end{thebibliography}
